\documentclass{article}
\usepackage{amssymb}
\usepackage{amsmath}

\usepackage[dvipsnames]{xcolor}
\usepackage{tikz}
\usetikzlibrary{decorations.markings}
\usetikzlibrary{positioning}
\tikzset{->-/.style={decoration={markings, mark=at position #1 with {\arrow{>}}},postaction={decorate}}}
\tikzset{-<-/.style={decoration={markings, mark=at position #1 with {\arrow{<}}},postaction={decorate}}}

\begin{document}

\title{Unbounded solutions of models for glycolysis}

\author{Pia Brechmann and Alan D. Rendall\\
Institut f\"ur Mathematik\\
Johannes Gutenberg-Universit\"at\\
Staudingerweg 9\\
D-55099 Mainz\\
Germany}

\date{}

\maketitle

\begin{abstract}
The Selkov oscillator, a simple description of glycolysis, is a system of 
two ordinary differential equations with mass action kinetics. In previous
work the authors established several properties of the solutions of this 
system. In the present paper we extend this to prove that this system has 
solutions which diverge to infinity in an oscillatory manner at late times. 
This system was originally derived from another system with Michaelis-Menten
kinetics. It is shown that the Michaelis-Menten system, like that with mass
action, has solutions which diverge to infinity in a monotone manner. It is
also shown to admit subcritical Hopf bifurcations and thus unstable periodic
solutions. We discuss to what extent the unbounded solutions cast doubt on the 
biological relevance of the Selkov oscillator and compare it with other models 
in the literature. 
\end{abstract}

\section{Introduction}

When trying to understand a biological system with the help of mathematical 
modelling it often happens that there are several different models for 
the same biological situation in the literature. In view of this it is 
important to have criteria for deciding between models. One strategy for 
identifying criteria of this type is to look at relatively simple examples in 
great detail. In order to do this effectively it is necessary to 
have a sufficiently comprehensive understanding of the properties of 
solutions of the models being studied. In this paper, with this strategy in 
mind, we look in detail at the dynamical properties of certain models for 
glycolysis.

Glycolysis is part of the process by which living organisms extract energy 
from sugar \cite{alberts08}. A suitable model system for studying this
phenomenon 
experimentally is yeast extract or a suspension of yeast cells. The first
indication that this system might have interesting dynamical properties was
given by damped oscillations reported in \cite{duysens57}. Later it was 
discovered that a constant continuous supply of sugar can lead to sustained 
oscillations (cf. \cite{boiteux75}). Looking for the source of these 
oscillations revealed that they are produced by a small reaction network 
describing the action of the enzyme phosphofructokinase. A mathematical model
for this network was set up and studied by Higgins \cite{higgins64}. It 
was found by Selkov that this model was not adequate for describing the 
oscillations and he introduced a modified one \cite{selkov68}. The starting
point for the model of Selkov is a reaction network with five chemical species. 
Assuming mass action kinetics leads to a system of five ordinary differential 
equations. Using quasi-steady state assumptions this can be reduced to a 
system of two equations with nonlinearities of Michaelis-Menten type. For 
brevity we call it 'the Michaelis-Menten system' in what follows. Setting one 
of the coefficients in this system to zero leads to a further simplification, 
giving a system of two equations with mass action kinetics, which we call the 
'basic Selkov system' in what follows. 

The aim of this paper is to obtain a better understanding of the dynamics of 
solutions of the three systems just described. A number of properties of 
solutions of the basic Selkov system were 
already established in \cite{selkov68} but for many years no further rigorous 
results on this subject were obtained. Important progress was made in a paper 
of d'Onofrio \cite{donofrio10} and a number of additional properties of the 
solutions were established in a recent paper of the authors 
\cite{brechmann18}. In particular it was proved that for any values of the 
parameters there exist unbounded solutions of this system which are 
eventually monotone in the sense that for a solution of this type 
both concentrations are monotone after a certain time. In \cite{selkov68} 
it is claimed that this system admits solutions which oscillate with an 
amplitude which grows without limit at late times. In what follows solutions 
of this type are referred to as 'solutions with unbounded oscillations'.
The paper \cite{selkov68} provides no justification for the claim other than a 
mention of numerical simulations, about which no details are given. Up to now 
there was no proof of the truth or falsity of this claim of \cite{selkov68}. 
One of the main results of the present paper is a proof of the existence of 
solutions of the basic Selkov system with unbounded oscillations. Our discovery 
of this proof was stimulated by the paper \cite{merkin87}, which belongs to 
the domain of theoretical chemistry. It deals with a system which turns out to 
be identical to the basic Selkov system when a parameter $\gamma$ in the latter
system takes the value two. 

In \cite{merkin87} a claim of the existence of solutions with unbounded 
oscillations is also made. It is supported by an intricate heuristic argument 
using matched asymptotic expansions. It is not at all clear how this argument 
could be translated into a rigorous one but it provided us with some ideas 
which, when combined with the results of \cite{brechmann18}, do give a proof
of the existence of solutions with unbounded oscillations. 
When written in dimensionless form the system contains one parameter $\alpha$. 
As claimed in \cite{merkin87}, solutions with unbounded oscillations occur 
for precisely one value $\alpha_1$ of $\alpha$. When $\alpha$ is slightly less
than $\alpha_1$ there exists a stable periodic solution. As $\alpha$ approaches 
$\alpha_1$ from below the amplitude of the periodic solution tends to 
infinity. One important element of this proof is to study the limit of the 
system for $\alpha\to\infty$ after a suitable rescaling. The existence of 
$\alpha_1$ is then proved by a shooting argument. A monotonicity property, 
which was apparently not previously known, is used to obtain the uniqueness of 
$\alpha_1$. 

The presence of unbounded solutions, whether monotone or oscillatory, might be
seen as a feature which is unrealistic from the point of view of the 
biological applications. The monotone unbounded solutions of the basic Selkov 
system are not mentioned at all in \cite{selkov68}. That system is the limit of 
the Michaelis-Menten system when a parameter $\nu$ tends to zero. It is stated 
in \cite{selkov68} that solutions with unbounded oscillations do not exist
for $\nu>0$. On the other hand simulations reported in \cite{keener09} 
suggest that the amplitude of periodic solutions of the Michaelis-Menten system
diverges rapidly to infinity when a parameter is varied in a finite range.
This indicates that, in contrast to the claim of Selkov, the existence of 
unbounded oscillations is a phenomenon which may persist for $\nu>0$. If this
is true then the presence of these biologically problematic solutions of the 
basic Selkov system is not just an artefact of taking the limit $\nu\to 0$. 
The issue of the existence of solutions with unbounded oscillations in the 
case of the Michaelis-Menten system is not resolved in what follows but some 
partial results are obtained. In particular it is shown that for the 
Michaelis-Menten system with arbitrary parameters there are unbounded 
solutions which are eventually monotone and whose leading order asymptotics 
are identical to those found in the basic Selkov system. It is also shown that 
for certain combinations of the parameters $(\alpha,\nu)$ with $\nu>0$ all 
positive solutions except the steady state have these late-time asymptotics. 
It turns out that there are parameter values for which there exist unstable 
periodic solutions of the Michaelis-Menten system. This is in contrast to the 
basic Selkov system where it was proved in \cite{brechmann18} that all periodic 
solutions are asymptotically stable. 

The structure of the paper is as follows. The various systems considered in 
the paper are defined in section \ref{survey}. In section \ref{basic},
after some necessary results on the basic Selkov system proved in 
\cite{brechmann18} have been recalled, the existence of solutions with 
unbounded oscillations is proved. Similarities and differences
between the properties of solutions of the basic Selkov system and the 
Michaelis-Menten system are 
discussed in the next three sections. Section \ref{mm} discusses the Hopf
bifurcation exhibited by the Michaelis-Menten system. Its Poincar\'e 
compactification is computed in section \ref{poincare}. Global properties of 
the Michaelis-Menten system are
discussed in section \ref{portrait}. The paper ends with a conclusion and
outlook. 

\section{Survey of the systems considered}\label{survey}
In \cite{selkov68} a simple reaction network describing glycolysis is 
introduced. Assuming mass action kinetics for this network leads to a 
system of five ordinary differential equations, system (4) of \cite{selkov68}.
In a slightly modified notation this system is
\begin{eqnarray}
&&\frac{ds_1}{dt}=v_1-k_1s_1x_1+k_{-1}x_2,\\
&&\frac{ds_2}{dt}=k_2x_2-\gamma k_3s_2^\gamma e+\gamma k_{-3}x_1-v_2s_2,\\
&&\frac{dx_1}{dt}=-k_1s_1x_1+(k_{-1}+k_2)x_2+k_3s_2^\gamma e-k_{-3}x_1,\\
&&\frac{dx_2}{dt}=k_1s_1x_1-(k_{-1}+k_2)x_2,\\
&&\frac{de}{dt}=-k_3s_2^\gamma e+k_{-3}x_1.
\end{eqnarray}
In fact  a factor $\gamma$ was omitted in two places in \cite{selkov68} and 
this error has been corrected here. All the parameters are positive and it is 
assumed that $\gamma>1$, which encodes the biological property of cooperativity.
Note that $e_0=e+x_1+x_2$ is a conserved quantity (total amount of enzyme) and 
this can be used to eliminate $e$ from the first four evolution equations and 
discard the evolution equation for $e$. This reduces the system to four 
equations.

Dimensionless variables can be introduced by defining 
\begin{equation}
\sigma_1=\frac{k_1s_1}{k_{-1}+k_2},\ 
\sigma_2=\left(\frac{k_3}{k_{-3}}\right)^{\frac{1}{\gamma}}s_2,\ 
u_1=\frac{x_1}{e_0},\ u_2=\frac{x_2}{e_0},\ 
\theta=\frac{e_0k_1k_2}{k_{-1}+k_2}t.
\end{equation}
This leads to the system
\begin{eqnarray}
&&\frac{d\sigma_1}{d\theta}=\nu-\frac{k_2+k_{-1}}{k_2}u_1\sigma_1
+\frac{k_{-1}}{k_2}u_2,\label{selkove1}\\
&&\frac{d\sigma_2}{d\theta}=\eta\left(u_2-\gamma\frac{k_{-3}}{k_2}
\sigma_2^\gamma(1-u_1-u_2)+\gamma\frac{k_{-3}}{k_2}u_1-\chi\sigma_2\right),
\label{selkove2}\\
&&\epsilon\frac{du_1}{d\theta}=u_2-\sigma_1u_1+\frac{k_{-3}}{k_2+k_{-1}}
(\sigma_2^\gamma(1-u_1-u_2)-u_1),\label{selkove3}\\
&&\epsilon\frac{du_2}{d\theta}=\sigma_1u_1-u_2\label{selkove4}
\end{eqnarray}
where 
\begin{equation}
\epsilon=\frac{e_0k_1k_2}{(k_2+k_{-1})^2},\ \nu=\frac{v_1}{k_2e_0},\ 
\eta=\frac{k_2+k_{-1}}{k_1}\left(\frac{k_3}{k_{-3}}\right)^{\frac{1}{\gamma}},\
\chi=\frac{v_2}{k_2e_0}\left(\frac{k_{-3}}{k_3}\right)^{\frac{1}{\gamma}}.
\end{equation}
Formally setting 
$\epsilon=0$ in the equations (\ref{selkove3}) and (\ref{selkove4}) gives 
$u_2=\sigma_1 u_1$ and 
$u_1=\frac{\sigma_2^\gamma}{1+\sigma_2^\gamma+\sigma_1\sigma_2^\gamma}$ and
substituting these relations into the evolution equations for $\sigma_1$ and 
$\sigma_2$ gives
\begin{eqnarray}
&&\frac{d\sigma_1}{d\theta}=\nu-\left(
\frac{\sigma_1\sigma_2^\gamma}{1+\sigma_2^\gamma+\sigma_1\sigma_2^\gamma}\right),
\label{selkovlim1}\\
&&\frac{d\sigma_2}{d\theta}=\eta \left(
\frac{\sigma_1\sigma_2^\gamma}{1+\sigma_2^\gamma+\sigma_1\sigma_2^\gamma}
-\chi\sigma_2\right).\label{selkovlim2}
\end{eqnarray}
As has been discussed in \cite{brechmann18} geometric singular perturbation 
theory (GSPT) can
be used to show that solutions of (\ref{selkove1})-(\ref{selkove4})
converge to solutions of (\ref{selkovlim1})-(\ref{selkovlim2}) in the limit 
$\epsilon\to 0$.

In \cite{selkov68} a further simplification of this system is introduced.
Consider the rescaled quantities
\begin{equation}
x=\frac{\nu^{\gamma-1}}{\chi^\gamma}\sigma_1,\ y=\frac{\chi}{\nu}\sigma_2,\ 
\alpha=\frac{\eta\chi^{\gamma+1}}{\nu^\gamma},\ 
\beta=\frac{\nu^{\gamma-1}}{\chi^\gamma}, 
\tau=\left(\frac{\nu}{\chi}\right)^\gamma\theta.
\end{equation}
Expressing the equations (\ref{selkovlim1}) and (\ref{selkovlim2}) in terms of 
these gives
\begin{eqnarray}
&&\frac{dx}{d\tau}=1-\frac{xy^\gamma}{1+\nu y^\gamma(\beta+x)}\label{selkovmm1},\\
&&\frac{dy}{d\tau}=\alpha\left[\frac{xy^\gamma}{1+\nu y^\gamma(\beta+x)}-y\right].
\label{selkovmm2}
\end{eqnarray}
This system has a regular limit when $\nu$ tends to zero with $\alpha$ and 
$\beta$ fixed. In the limit we get the basic Selkov system, system (II) of 
\cite{selkov68}, which is 
\begin{eqnarray}
&&\frac{dx}{d\tau}=1-xy^\gamma,\label{selkov1}\\
&&\frac{dy}{d\tau}=\alpha y(xy^{\gamma-1}-1).\label{selkov2}
\end{eqnarray}
It is the system of central interest in \cite{selkov68} and the dynamical
properties of its solutions are studied in detail in \cite{brechmann18}.
Of course (\ref{selkov1})-(\ref{selkov2}) can be thought of as the special case
of (\ref{selkovmm1})-(\ref{selkovmm2}) where $\nu=0$.

\section{The basic Selkov system}\label{basic}

The following proposition collects some of the properties of solutions of the
basic Selkov system established in \cite{brechmann18}.

\noindent
{\bf Proposition 1} The basic Selkov system (\ref{selkov1})-(\ref{selkov2}) has 
the following properties.

\noindent
1. For each value of the parameter $\alpha$ the unique positive steady state has
coordinates $(1,1)$.

\noindent
2. For each $\alpha\in \left(0,\frac{1}{\gamma-1}\right)$ the positive steady 
state is asymptotically stable and there exist no periodic solutions.

\noindent
3. For $\alpha=\alpha_0=\frac{1}{\gamma-1}$ a generic supercritical Hopf 
bifurcation occurs.

\noindent
4. For each value of the parameter $\alpha$ there exist positive numbers
$x_0$ and $y_0$ such that if a solution satisfies $x(t)\ge x_0$ and 
$y(t)\le y_0$ at some time $t$ it satisfies $\dot x(t)>0$ and $\dot y(t)<0$ at
all later times, $\lim_{t\to\infty}x(t)=\infty$ and $\lim_{t\to\infty}y(t)=0$.

Using the standard theory of Hopf bifurcations it follows from statement 3. of
the proposition that for any $\alpha$ slightly greater than $\alpha_0$ there
exists a stable periodic solution. A key question left open in 
\cite{brechmann18} is that of what happens to the periodic solution when
$\alpha$ gets large. This question is answered in this section.

\noindent
{\bf Theorem 1} There exists a number $\alpha_1>\alpha_0$ such that the basic 
Selkov system (\ref{selkov1})-(\ref{selkov2}) has the following properties.

\noindent
1. For $\alpha=\alpha_1$ there exist solutions with the properties that
$\liminf_{t\to\infty}x(t)=\liminf_{t\to\infty}y(t)=0$ and
$\limsup_{t\to\infty}x(t)=\limsup_{t\to\infty}y(t)=\infty$.

\noindent
2. For $\alpha_0<\alpha<\alpha_1$ there exists a unique periodic solution and 
it is asymptotically stable.

\noindent 
3. For $\alpha>\alpha_1$ each solution other than the steady state is unbounded
and has the properties described in statement 4. of Proposition 1. 

\noindent
4. As $\alpha$ tends to $\alpha_1$ from below the diameter of the image of the 
periodic solution tends to infinity. 

We adopt some of the notation of \cite{brechmann18}. There the Poincar\'e 
compactification of the basic Selkov system is computed and one of the
resulting points at infinity is blown up. After this has been done there are
four steady states at infinity called $P_1$, $P_2$, $P_3$ and $P_4$. Their
positions can be seen in Fig. 1 of \cite{brechmann18}. Each of the points
$P_1$ and $P_3$ has a one-dimensional centre manifold with the flow on the 
centre manifolds being away from $P_1$ and towards $P_3$. As a starting point 
for the proof of Theorem 1 we establish some further properties of the centre 
manifolds of the points $P_1$ and $P_3$, both of which are unique. Let $L$ be 
the segment of the line $y=1$ where $0<x\le 1$. We use the notation for the
components $U_i$ of the complement of the nullclines which can be seen in Fig. 2
of  \cite{brechmann18}.

\noindent
{\bf Lemma 1} In the basic Selkov system the centre manifolds of $P_1$ and $P_3$
both contain a point of $L$ in their closures.

\noindent
{\bf Proof} For a point on the centre manifold of $P_1$ sufficiently near to 
$P_1$ we have $\dot x>0$. Hence the manifold initially
lies in the region $U_1$. As long as $x<1$ it must remain in $U_1$ and both 
coordinates of a solution on the centre manifold are monotone functions of 
time. Hence a solution on the centre manifold of $P_1$ either reaches a 
point of $L$ with $x<1$ after a finite time or it tends to the positive steady 
state as $t\to\infty$. Similarly a solution on the centre manifold of $P_3$, 
when followed backwards in time, either reaches a point of $L$ with $x<1$ after 
a finite time or it tends to the positive steady state as $t\to -\infty$. 
$\blacksquare$

For a given value of $\alpha$ let $\xi_1(\alpha)$ be the $x$-coordinate of the 
point where the centre manifold of $P_1$ meets $L$ if such a point exists and 
otherwise let $\xi_1(\alpha)=1$. Define $\xi_2(\alpha)$ similarly in terms of 
the 
centre manifold of $P_3$. Note that each centre manifold depends smoothly on 
the parameter $\alpha$, in the sense that we can choose initial data for 
solutions on the centre manifold for different values of $\alpha$ in such a
way that the solutions depend smoothly on $\alpha$. This can be seen by 
considering the suspended system obtained by adjoining the equation
$\dot\alpha=0$ to the basic Selkov system and noting that it has a 
two-dimensional centre manifold at the points corresponding to $P_1$ and $P_3$.
This manifold is foliated by curves of constant $\alpha$ which are centre 
manifolds for the original system. Their smooth dependence on $\alpha$ follows 
from the smoothness of the two-dimensional centre manifold. 

\noindent
{\bf Lemma 2} The function $\xi_1-\xi_2$ describing the separation of the points
where the centre manifolds of $P_1$ and $P_3$ reach $y=1$ is continuous.

\noindent
{\bf Proof} Consider a value of $\alpha_c$ for which $\xi_1(\alpha_c)<1$.
The centre manifold for that value crosses $L$ transversely and so, by the
implicit function theorem, $\xi_1$ is a smooth function of $\alpha$ close to
$\alpha_c$. This also shows that the set of values of $\alpha$ for which 
$\xi_1(\alpha)<1$ is open. Consider now a value $\alpha^*$ of $\alpha$ for 
which $\xi_1(\alpha^*)=1$ and a sequence $\alpha_n$ satisfying 
$\lim_{n\to\infty}\alpha_n=\alpha^*$. It will be shown that 
$\lim_{n\to\infty}\xi_1(\alpha_n)=1$. Together with the information already 
obtained this implies that $\xi_1$ is continuous everywhere. The desired 
statement will be proved by contradiction. If $\xi_1(\alpha_n)$ did not 
converge to one then by passing to a subsequence we could assume that 
$\lim_{n\to\infty}\xi_1(\alpha_n)=\xi_s<1$. Consider now the sequence of 
solutions of the basic Selkov system with $x_n(0)=\xi_1(\alpha_n)$, 
$y_n(0)=1$ and $\alpha=\alpha_n$ and the solution with $x_s(0)=\xi_s$,
$y_s(0)=1$ and $\alpha=\alpha^*$. We are interested in these solutions for 
$t\le 0$. The sequence $(x_n,y_n)$ converges to $(x_s,y_s)$ uniformly on 
compact time intervals. We 
claim that $(x_s,y_s)$ lies on the centre manifold of $P_1$ for 
$\alpha=\alpha^*$. If $(x_s,y_s)$ lies to the left of the centre manifold then 
it reaches negative values of $x$ for finite negative values of $t$. Then for 
$n$ sufficiently large the solutions $(x_n,y_n)$ would do the same, a 
contradiction. If $(x_s,y_s)$ lies to the right of the centre manifold then it 
must reach values of $x$ greater than $\xi_s$ for finite negative 
values of $t$. Then for $n$ sufficiently large the solutions $(x_n,y_n)$ would 
do the same, a contradiction. The conclusion is that the solution
$(x_s,y_s)$ lies on the centre manifold and hence $\xi_1(\alpha^*)<1$, in 
contradiction to the definition of $\alpha^*$. It has 
thus been proved that $\xi_1$ is continuous. A similar argument shows that 
$\xi_2$ is continuous. Hence $\xi_1-\xi_2$ is continuous.  $\blacksquare$

\noindent
{\bf Lemma 3} The function $\xi_1-\xi_2$ describing the separation of the points
where the centre manifolds of $P_1$ and $P_3$ reach $y=1$ is positive for 
$0<\alpha\le\alpha_0$ and negative for $\alpha$ sufficiently large. There 
exists an $\alpha_1$ with $\xi_1(\alpha_1)= \xi_2(\alpha_1)$.

\noindent
{\bf Proof} Suppose that for a given value of $\alpha$ we have 
$(\xi_1-\xi_2)(\alpha)\le 0$. The region of the Poincar\'e compactification 
bounded by the parts of the centre manifolds of $P_1$ and $P_3$ ending on $L$ 
and the part of $L$ between them and above the centre manifold of $P_3$ is 
invariant under evolution backwards in time. Consider the solution obtained by
backward time evolution of a point in this region other than the steady state.
By Poincar\'e-Bendixson theory its $\alpha$-limit set must be a steady state
or a periodic solution. If $\alpha\le\alpha_0$ this leads to a contradiction,
because in that case no periodic solutions exist and the positive steady state
is a sink. Thus we can conclude that the function $\xi_1-\xi_2$ is positive
for $0<\alpha\le\alpha_0$.

Next we investigate the behaviour of solutions for $\alpha$ large. The following
calculations were inspired by a transformation introduced in \cite{merkin87}
in the case $\gamma=2$. It is given by $\mu=\alpha^{-\frac{1}{\gamma}}$,
$\tilde x=\alpha^{\frac{\gamma-1}{\gamma}}x$, $\tilde y=\alpha^{-\frac{1}{\gamma}}y$
and $\tilde\tau=\alpha\tau$. The equations become
\begin{eqnarray}
&&\frac{d\tilde x}{d\tilde\tau}=\mu-\tilde x\tilde y^\gamma,\label{merkin1}\\
&&\frac{d\tilde y}{d\tilde\tau}=\tilde x\tilde y^\gamma-\tilde y\label{merkin2}
\end{eqnarray} 
and we are interested in the limit $\mu\to 0$. A Poincar\'e compactification
of this system was carried out in \cite{merkin87}. After a suitable rescaling
this leads to a system in the standard form of a fast-slow system in GSPT.
(For background on GSPT we refer to \cite{kuehn15}.) Unfortunately in this
system the important propery of normal hyperbolicity breaks down at the
point corresponding to $P_3$. It turns out that this problem can be got around
by using the transformations introduced in \cite{brechmann18} to treat the
behaviour of solutions for $x$ large. These can be summed up by defining
$\bar y=x^{-\frac{1}{\gamma}}y^{\frac{1}{\gamma}}$ and 
$\bar z=x^{-\frac{1}{\gamma}}y^{-\frac{\gamma-1}{\gamma}}$ and choosing a time coordinate
$s$ satisfying $\frac{ds}{d\tau}=\frac{1}{\gamma}xy^{\gamma-1}$. This transforms 
the basic Selkov system into system (12)-(13) of \cite{brechmann18}. Now 
introduce $\epsilon=\alpha^{-1}$ and $\bar w=\alpha(\bar z-1)$. Then, denoting 
the derivative with respect to $s$ by a prime, we get the system
\begin{eqnarray}
&&\bar y'=-\gamma\bar y\bar w
-\bar y\epsilon^{-1}[(1+\epsilon\bar w)^\gamma-1-\gamma\epsilon\bar w]
+\bar y^{\gamma+1}\nonumber\\
&&-\bar y^\gamma(1+\epsilon\bar w)^{\gamma+1},\label{fastslow1}\\
&&\epsilon \bar w'=\gamma (\gamma-1)\bar w
-(\gamma-1)\bar y^\gamma(1+\epsilon\bar w)\nonumber\\
&&+(\gamma-1)\epsilon^{-1}[(1+\epsilon\bar w)^{\gamma+1}
-1-\epsilon(\gamma+1)\bar w]\nonumber\\
&&-\bar y^{\gamma-1}(1+\epsilon\bar w)^{\gamma+2}
+\gamma\bar y^\gamma(1+\epsilon\bar w).\label{fastslow2}
\end{eqnarray}
Note that, due to cancellations in the expressions in square brackets this
system is regular at $\epsilon=0$ and in fact the apparently singular term
even vanishes as $\epsilon\to 0$. The critical manifold has the equation 
$\gamma (\gamma-1)\bar w=\bar y^{\gamma-1}-\bar y^\gamma
=\bar y^{\gamma-1}(1-\bar y)$. 
The derivative of the right hand side of the equation (\ref{fastslow2})
with respect to $\bar w$, evaluated at $\epsilon=0$, is $\gamma (\gamma-1)$. 
Thus the critical manifold is normally hyperbolic repelling. (For the 
terminology see \cite{kuehn15}.) The evolution equation on the critical 
manifold is
\begin{equation}
\frac{d\bar y}{ds}=-\frac{\gamma}{\gamma-1}\bar y^\gamma(1-\bar y).
\end{equation}
On the critical manifold there are two steady states, a source and a 
sink. They are connected by a heteroclinic orbit. For $\epsilon$ small and
positive the critical manifold perturbs to a one-dimensional invariant 
manifold. All steady states which exist must lie on that manifold. The 
steady state at $\bar y=1$ is hyperbolic and so perturbs to a hyperbolic
source. The steady state at $\bar y=0$ continues to exist and there are no 
others. It follows that there is also a connection between the positive steady 
state of the Selkov system and the point $P_3$ on the boundary for $\alpha$ 
sufficiently large. In other words, when $\alpha$ is sufficiently large the 
centre manifold of $P_3$ converges to the positive steady state in the past.
This means that $\xi_2(\alpha)=1$. On the other hand, since the positive steady 
state is a source in this case the centre manifold of $P_1$ cannot converge to
the positive steady state. We conclude that $\xi_1(\alpha)<1$ and that 
$\xi_1-\xi_2$ is negative. By the intermediate value theorem there exists some 
$\alpha_1$ with $\xi_1(\alpha_1)=\xi_2(\alpha_1)$. Note that in the end the
equations (\ref{merkin1})-(\ref{merkin2}) were not needed in the proof but
we judged it useful to include them so as to give an indication of how the
argument was found.
$\blacksquare$

It turns out that the value of $\alpha$ for which the centre manifolds of $P_1$
and $P_3$ meet is unique. This follows from a monotonicity property of the 
dependence of the centre manifolds on $\alpha$.

\noindent
{\bf Lemma 4} The function $\xi_1-\xi_2$ describing the separation of the points
where the centre manifolds of $P_1$ and $P_3$ reach $y=1$ is strictly 
decreasing and has a unique zero.

\noindent
{\bf Proof} For this proof it is convenient to think of $y$ as a function of 
$x$ for a given solution. Suppose that a solution $y(x)$ 
for a parameter $\alpha$ crosses a solution $\hat y(x)$ for a parameter 
$\hat\alpha<\alpha$. Then the (negative) slope of $\hat y$ is smaller in 
magnitude than that of $y$. Thus if $\hat y$ is larger than $y$ for some $x$ 
it must remain so for all larger $x$. Similarly, if $\hat y$ is smaller than 
$y$ for some $x$ it must remain so for all smaller $x$. The leading order 
approximation to the centre manifold of $P_3$ is given by 
$\bar z=1+\nu_1\bar y^{\gamma-1}+\ldots$ where 
$\nu_1=\frac{1}{\alpha\gamma (\gamma-1)}$.
This translates (in terms of variables used in \cite{brechmann18})
to $Z=Y^{\frac{\gamma-1}{\gamma}}+\nu_1Y^{\frac{2(\gamma-1)}{\gamma}}\ldots$  and 
$x=y^{1-\gamma}-\gamma\nu_1+\ldots$. Putting these things together shows 
that when $\alpha$ is reduced the intersection of the centre manifold of $P_3$ 
with the line $y=1$ moves to the left. To obtain information about the 
position of the centre manifold of $P_1$ in its dependence on $\alpha$ it is 
necessary to determine one more order in the expansion of the centre manifold 
than was done in \cite{brechmann18}. The result is
$X=Z^{\gamma+1}-\gamma\alpha Z^{2\gamma+1}+\ldots$. In the original
variables this gives $x=y^{-\gamma}-\gamma\alpha y^{-2\gamma}+\ldots$.
When $\alpha$ is reduced $x$ 
becomes larger for fixed $y$. This also means that $y$ becomes larger for fixed
$x$ and this propagates to larger values of $x$. Thus the intersection of the 
centre manifold of $P_1$ with the line $y=1$ moves to the right. This implies
that the function $\xi_1-\xi_2$ is strictly decreasing and cannot have more 
than one zero.  $\blacksquare$

\noindent
{\bf Proof of Theorem 1} By Lemma 3 and Lemma 4 there exists a unique 
$\alpha_1>\alpha_0$ 
for which the centre manifolds of $P_1$ and $P_3$ coincide. With this 
information the first statement of Theorem 1 follows immediately from the 
first statement of Theorem 3 of \cite{brechmann18}. For 
$\alpha_0<\alpha<\alpha_1$ the positive steady state is unstable and there
is no heteroclinic cycle at infinity. It follows from the Poincar\'e-Bendixson
theorem that the $\omega$-limit set of a solution which starts near the steady
state but is not the steady state itself must be a periodic solution. In
particular, a periodic solution exists and we are in the second case of
Theorem 3 of \cite{brechmann18}. Thus the second statement of Theorem 1 holds.
If $\alpha>\alpha_1$ then there is again no heteroclinic cycle at infinity.
The $\alpha$-limit set of the solution on the centre manifold of $P_3$ must
then, by the Poincar\'e-Bendixson theorem, be either a periodic solution or the 
positive steady state. Moreover, if a periodic solution exists then only the
first possibility can occur. Since, however, it follows from \cite{brechmann18}
that any periodic solution which exists is stable the first possibility is 
ruled out. There can be no periodic solution and the third case of
Theorem 3 of \cite{brechmann18} must be realised. This completes the proof of
the third statement. Finally, the fourth statement will be proved by 
contradiction. Let $\beta_i$ be a sequence tending to $\alpha_1$ from below.
For a given $i$ the system with parameter $\beta_i$ has a unique
periodic solution and there is a unique point in its image of the form
$(1,z_i)$ with $z_i>1$. If this sequence did not tend to infinity then it 
would have a convergent subsequence. Thus after passing to a subsequence
$z_i$ tends to a finite limit $z^*$. The periodic solutions through the 
points $(1,z_i)$ converge to a solution through the point $(1,z^*)$, which is
a periodic solution of the system with parameter value $\alpha_1$. This
contradicts the fact that there are no such solutions.  $\blacksquare$   

\section{The Michaelis-Menten system}\label{mm}

In the system (\ref{selkovmm1})-(\ref{selkovmm2}) the $x$-axis is an invariant 
manifold of the flow and the vector field is directed toward positive values 
of $x$ on the $y$-axis. For each fixed choice of the parameters with $\nu<1$ 
there is a unique positive steady state at 
$\left(\frac{1+\beta\nu}{1-\nu},1\right)$. For $\nu\ge 1$ there is
no positive steady state. Linearizing the system about the steady state 
leads to the Jacobian
\begin{equation}
J=\left[
{\begin{array}{cc}
-\frac{(1-\nu)^2}{1+\beta\nu} & -\gamma\left(\frac{1-\nu}{1+\beta\nu}\right)\\
\alpha\frac{(1-\nu)^2}{1+\beta\nu} &
\alpha \left(\gamma\left(\frac{1-\nu}{1+\beta\nu}\right) -1\right)\\
\end{array}}
\right].
\end{equation}
The determinant of $J$ is $\alpha\frac{(1-\nu)^2}{1+\beta\nu}$ which is 
always positive. Thus the stability of the steady state is determined by the
trace of $J$, which is 
\begin{equation}
\alpha\left[\gamma\left(\frac{1-\nu}{1+\beta\nu}\right)-1\right]
-\frac{(1-\nu)^2}{1+\beta\nu}.
\end{equation}
If $\gamma\le\frac{1+\beta\nu}{1-\nu}$ then the trace of $J$ is negative for 
all values of $\alpha$ and the steady state is always stable. If 
$\gamma>\frac{1+\beta\nu}{1-\nu}$ define 
$\alpha_0=\frac{(1-\nu)^2}{\gamma (1-\nu)-(1+\beta\nu)}$. 
Then for $\alpha<\alpha_0$ the 
trace of $J$ is negative and the steady state is asymptotically stable while
for $\alpha>\alpha_0$ the trace of $J$ is positive and the steady state is a 
source. For $\alpha=\alpha_0$ there is a pair of imaginary eigenvalues. If we
consider the real part of the eigenvalues as a function of $\alpha$ then it 
passes through zero when $\alpha=\alpha_0$ and its derivative with respect to 
$\alpha$ at that point is non-zero. Thus a Hopf bifurcation occurs.

In the limiting case $\nu=0$ it was shown in \cite{brechmann18} that the Hopf
bifurcation is supercritical so that there exists a stable periodic solution for
any $\alpha$ slightly greater than $\alpha_0$. The computation of the 
Lyapunov number required to obtain this conclusion becomes considerably more
complicated for $\nu>0$. Rather than trying to do this in general we will
confine ourselves to obtaining some information for restricted sets of 
parameters. The Lyapunov number of the Hopf bifurcation is a function of
the parameters $\alpha$, $\beta$, $\gamma$ and $\nu$ and we are interested 
in its sign. A general formula for this quantity is given in Section 4.4 of
\cite{perko01}. It is of the form $\frac{-3\pi}{2b\Delta^{3/2}}f$, where
the first factor is positive in the present case and $f$ is a function of
$(\alpha,\beta,\gamma,\nu)$ which is negative when $\nu=0$. This shows 
that in that case the Hopf bifurcation is supercritical. For $\nu$ small
and positive $f$ is still negative and the bifurcation supercritical. It will 
now be proved that there also exist parameters for which $f$ is positive, so 
that there exists a subcritical Hopf bifurcation. In that case there exist 
unstable periodic solutions for $\alpha$ slightly less than $\alpha_0$. Note 
for comparison that it was shown in \cite{brechmann18} that for $\nu=0$ 
unstable periodic solutions never exist. It suffices to treat the case $\beta=0$
since an example with $\beta$ small and positive follows by continuity. Since
we are only looking for some example we can also restrict to the case 
$\gamma=2$. 

With a suitable normalization the function $f$ is of the following form.
\begin{eqnarray}
&&\alpha (1-\nu)^2(-a_{11}^2+2\alpha a_{11}a_{02})\nonumber\\
&&+2(1-\nu)(\alpha^2a_{11}^2-\alpha a_{11}(a_{02}+a_{20}))\nonumber\\
&&+\alpha^2(1-\nu)^2(a_{11}a_{02}-2\alpha a_{02}^2)
+2\alpha (1-\nu)^2(\alpha^2a_{02}^2-a_{20}a_{02})\nonumber\\
&&+4(1-\nu)(-a_{20}^2+\alpha^2a_{20}a_{02})
+4(2\alpha a_{20}^2-\alpha^2a_{11}a_{20})\\
&&+(2\alpha (1-\nu)+2(1-\nu)^2)(-\alpha^2a_{11}a_{02}+a_{11}a_{20})
+(1-\nu)^2[2\alpha-(1-\nu)]\times\nonumber\\
&&[3(-\alpha^2(1-\nu)a_{03}+2 a_{30})+2(1-\nu)(-a_{21}+\alpha a_{12})
+\alpha((1-\nu)a_{12}-2a_{21})]\nonumber
\end{eqnarray}
Here the notation $a_{ij}$ is taken from Perko \cite{perko01}. In order 
that there exist a bifurcation a restriction on $\nu$ must be satisfied and in 
the case $\gamma=2$ it is given by $\nu<\frac12$. Consider now the limit 
$\nu\to\frac12$. Since $\alpha=\frac{(1-\nu)^2}{1-2\nu}$ at the bifurcation 
point it tends to infinity in this limit. The highest power of $\alpha$ in the 
above expression is $\alpha^3$ and two terms containing $\alpha^3$ cancel.
Substituting in the expression for the bifurcation point gives a function
depending on $\nu$ alone and we want to examine its behaviour near
$\nu=\frac12$. To do this it suffices to retain only those terms in the above
expression which contain a power of $\alpha$ which is at least two. It is also
the case that the expressions for $a_{11}$ and $a_{03}$ contain a factor of
$1-2\nu$. Thus to order $(1-2\nu)^{-2}$ we get the expression
\begin{equation}
\frac14\alpha^2[-4(\alpha a_{11})a_{02}-3(\alpha a_{03})+8a_{20}a_{02}
+3a_{12}-4a_{21}]+\ldots
\end{equation}
The expression in square brackets tends to a positive value as $\nu\to\frac12$.
Thus the leading term in the expression for the Lyapunov number is positive
for $\nu$ close to its limiting value. This proves the desired statement.

\section{The Poincar\'e compactification}\label{poincare}

In \cite{brechmann18} it was investigated using the Poincar\'e
compactification in which ways solutions of (\ref{selkov1})-(\ref{selkov2})
can tend to infinity for large times. Here we want to carry out  
corresponding calculations for (\ref{selkovmm1})-(\ref{selkovmm2}). A useful
preliminary step is to introduce a new time coordinate $T$ satisfying
$\frac{d\tau}{dT}=1+\nu y^\gamma(\beta+x)$. Then we get the system
\begin{eqnarray}
&&\frac{dx}{dT}=1-xy^\gamma+\nu y^\gamma(\beta+x),\\
&&\frac{dy}{dT}=\alpha[xy^\gamma-y-\nu y^{\gamma+1}(\beta+x)].
\end{eqnarray}
This makes the right hand side into a polynomial while leaving the phase 
portrait unchanged. 

\begin{figure}
\centering
\begin{tikzpicture}[scale=1]
	\draw[thick,->-=.25,-<-=.75] (-4,3) arc(-90:-8:1cm and 1cm) coordinate[name=p2,midway] coordinate[name=p3];
	\draw[thick,->-=.5] (-4,0)--(-1.65,0) coordinate[name=p8];
	\draw[thick,-<-=.25,->-=.75] (p8) arc(180:100:1cm and 1cm) coordinate[name=p7,midway] coordinate[name=p6];
	\draw[thick,->-=.25,-<-=.75] (p6) arc(170:80:0.5cm and 0.5cm) coordinate[name=p5,midway] coordinate[name=p4]; 
	\draw[thick,-<-=.5] (p4) to [out=112,in=-15] (p3);
	\draw[thick,->] (-4.15,2)--(-3.85,2);
	
	\coordinate[below right=0.25cm and 0.25cm of p2] (z2); 
	\draw[thick,->-=.6,ForestGreen] (p2) to (z2);
	\draw[thick,->-=.6,ForestGreen,densely dotted] (p3) arc(-8:-25:1cm and 1cm);
	\coordinate[above left=0.25cm and 0.25cm of p5] (z5);
	\draw[thick,->-=.6,ForestGreen] (p5) to (z5);
	\coordinate[above left=0.25cm and 0.25cm of p7] (z7);
	\draw[thick,-<-=.5,ForestGreen] (p7) to (z7);
	\coordinate[left=0.3cm of p8] (z8);
	\draw[thick,-<-=.6,ForestGreen,densely dotted] (p8) to (z8);
	
	\node[gray] (g1) at (-4,4) {\small $\gamma_1$};
	\node[gray] (g4) at (-0.6,0) {\small $\gamma_4$};
	\node[gray] (g3) at (0,0.6) {\small $\gamma_3$};
	\node[gray] (g2) at (-1.5,3.6) {\small $\gamma_2$};
	
	\draw[thick] (-4,0)--(-4,3) node[above left]{\small $P_1$};
	\draw[fill] (p2) circle(1pt) node[above left]{\small $P_2$};
	\draw[fill] (p3) circle(1pt) node[above left]{\small $P_3$};
	\draw[fill] (p8) circle(1pt) node[below right]{\small $P_8$};
	\draw[fill] (p7) circle(1pt) node[below right]{\small $P_7$};
	\draw[fill] (p6) circle(1pt) node[below right]{\small $P_6$};
	\draw[fill] (p5) circle(1pt) node[below right]{\small $P_5$};
	\draw[fill] (p4) circle(1pt) node[below right]{\small $P_4$};
\end{tikzpicture}
\caption{Poincar\'e compactification.}
\label{FigPoincare}
\end{figure}
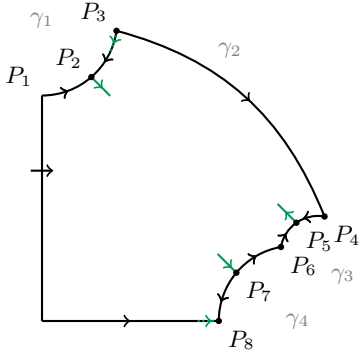

The phase portrait is more complicated than that in the case of mass action
kinetics. A schematic picture of it is given in Fig.~\ref{FigPoincare} and its properties are
summarized in the following lemma which is the 
analogue of Lemma 2 in \cite{brechmann18}.

\noindent
{\bf Lemma 5} Suppose that $\nu<1$. There is a smooth mapping of the closure 
of the positive quadrant into itself mapping the axes into themselves with the 
following properties. The restriction of $\phi$ to the open quadrant is a 
diffeomorphism onto its image. This image is a region whose closure is a 
compact set bounded by intervals $[0,x_0]$ and $[0,y_0]$ on the $x$- and 
$y$-axes and four smooth curves $\gamma_i, 1\le i\le 4$. The curve $\gamma_1$ 
joins the point $P_1=(0,y_0)$ with
a point $P_3$ in the positive quadrant. $\gamma_2$ joins the point $P_3$ with
the point $P_4$. For $3\le i\le 4$ the curve $\gamma_i$ joins the point 
$P_{2i-2}$ with the point $P_{2i}$ and $P_8=(x_0,0)$. The image of the 
dynamical system can be rescaled so as to extend smoothly to the closure of the
image of $\phi$ in such a way that $P_3$ and $P_{2i}, 2\le i\le 4$, are 
steady states and the $\gamma_i$ and the image of the $x$-axis under $\phi$
are invariant manifolds. There are further steady states $P_2$ and 
$P_{2i+1}, 2\le i\le 3$, on the boundary belonging to the interior of 
$\gamma_1$ and $\gamma_{i+1}, 2\le i\le 3$, respectively.

To analyse the case where $x$ becomes large (Case 1 in the terminology of 
\cite{brechmann18}) introduce the variables $Y=\frac{y}{x}$, $Z=\frac{1}{x}$. 
Define a new time variable $t$ satisfying $\frac{dt}{dT}=Z^{-\gamma-1}$. The 
result of the transformation is
\begin{eqnarray}
&&\frac{dY}{dt}=\alpha Y^\gamma Z+Y^{\gamma+1}Z-\alpha YZ^{\gamma+1}-YZ^{\gamma+2}
\nonumber\\
&&-\nu Y^{\gamma+1}(\alpha+Z)(1+\beta Z),\\
&&\frac{dZ}{dt}=Y^\gamma Z^2-Z^{\gamma+3}-\nu Y^\gamma Z^2(1+\beta Z).
\end{eqnarray}
Both axes are invariant under the flow and the flow there is towards the origin.
The linearization of the system about
the origin is identically zero. Thus we do a quasihomogeneous directional 
blow-up. An appropriate transformation can be obtained by using a Newton 
polygon as in \cite{dumortier06}. The coefficients 
are $\alpha=\gamma$ and $\beta=\gamma-1$. (These are the same values
as occurred in the blow-up of the corresponding point for the model
(\ref{selkov1})-(\ref{selkov2}).) Thus we use variables $\bar y$ and 
$\bar z$ satisfying $(Y,Z)=(\bar y^\gamma,\bar y^{\gamma-1}\bar z)$. In addition
we introduce a new time coordinate $s$ satisfying 
$\frac{ds}{dt}=\gamma^{-1}\bar y^{\gamma^2-1}$. The system becomes
\begin{eqnarray}
&&\frac{d\bar y}{ds}=\alpha\bar y\bar z+\bar y^{\gamma+1}\bar z
-\alpha\bar y\bar z^{\gamma+1}-\bar y^\gamma\bar z^{\gamma+2}
\nonumber\\
&&-\nu(\alpha+\bar y^{\gamma-1}\bar z)\bar y^2(1+\beta\bar y^{\gamma-1}\bar z),\\
&&\frac{d\bar z}{ds}=-\alpha(\gamma-1)\bar z^2+\bar y^{\gamma}\bar z^2
+\alpha(\gamma-1)\bar z^{\gamma+2}-\bar y^{\gamma-1}\bar z^{\gamma+3}
\nonumber\\
&&+\nu[(\gamma-1)\alpha-\bar y^{\gamma-1}\bar z]\bar y\bar z 
(1+\beta\bar y^{\gamma-1}\bar z).
\end{eqnarray}
Both axes are invariant under the flow. There is a steady state at the origin 
and one at the point $(0,1)$, which corresponds to $P_7$. The linearization at 
the origin is identically zero. 

Next the centre manifold of $P_7$ will be studied. Introducing $w=\bar z-1$ 
moves the steady state to origin. The centre subspace is given $w=\rho\bar y$ 
with $\rho=\frac{1-\alpha\nu}{2\alpha}$ for $\gamma=2$ and 
$\rho=-\frac{\nu}{\gamma}$ for $\gamma>2$. Consider now the case $\gamma=2$,
where
\begin{equation}
{\bar y}'=\alpha\bar y\bar z-\alpha\bar y\bar z^{\gamma+1}
-\alpha\nu\bar y^2-\bar y^\gamma+O(\bar y^{\gamma+1}).
\end{equation}
Substituting the asymptotic expansion for $\bar z$ in terms of $\bar y$ which
holds on the centre manifold into this relation gives
\begin{equation}
{\bar y}'=\alpha\rho\bar y^2-3\alpha\rho\bar y^2-\alpha\nu\bar y^2-\bar y^2
+O(\bar y^3)=-2\bar y^2+O(\bar y^3).
\end{equation}

\noindent
{\bf Lemma 6} For $\gamma>2$ the relation 
${\bar y}'=-\frac{\gamma}{\gamma-1}\bar y^\gamma+o(\bar y^\gamma)$ holds on the
centre manifold of $P_7$.

\noindent
{\bf Proof}  
We use the relation 
\begin{equation}{\bar z}'=-\bar y^{\gamma-1}
+(\gamma-1)[-\alpha\bar z^2+\alpha\bar z^{\gamma+2}
+\alpha\nu\bar y\bar z]+O(\bar y^\gamma).
\end{equation}
Substituting this into the evolution equation for $\bar y$ gives
\begin{equation}
{\bar y}'=-\frac{\gamma}{\gamma-1}\bar y^{\gamma}
-\frac{1}{\gamma-1}\bar y\bar z^{-1}{\bar z}'+O(\bar y^{\gamma+1}).
\end{equation}
With this it is possible to adapt the argument used to analyse the centre 
manifold of $P_3$ in \cite{brechmann18} to get the desired conclusion as
follows. Suppose that we know that ${\bar y}'=O(\bar y^k)$ for some $k$ 
with $2\le k\le\gamma-1$. Then it follows that ${\bar z}'=O(\bar y^{k+1})$.
Hence ${\bar y}'=O(\bar y^{k+1})$. After finitely many steps we get the 
second conclusion of the lemma. $\blacksquare$

We see that the flow on the centre manifold is towards $P_7$ and since the 
non-zero eigenvalue of the linearization at that point is positive $P_7$ is
a topological saddle. In fact the flow on the boundary is everywhere away
from $P_7$. We next blow up the origin in the coordinates 
$(\bar y,\bar z)$. This time the procedure described in 
\cite{dumortier06} leads to the choice of coefficients $\alpha=\beta=1$.
Blow-ups in the two coordinate directions are required. The only terms in the
resulting equations which will be written explicitly are those which have a
direct influence on the analysis which follows. In the first transformed 
system, with $\bar y=\tilde y_1$ and $\bar z=\tilde y_1\tilde z_1$, the 
equations are 
\begin{eqnarray}
  &&{\tilde y_1}'=\tilde y_1[-\alpha(\nu-\tilde z_1)\tilde y_1+\cdots],
     \label{trans1a}\\
&&{\tilde z_1}'=\tilde y_1[\gamma\alpha(\nu-\tilde z_1)\tilde z_1+\cdots].
\label{trans1b}
\end{eqnarray}
A change of time coordinate eliminates the common factor $\tilde y_1$.
On the boundary there is a steady state at the point $(0,\nu)$, which 
corresponds to $P_5$. The origin of this coordinate system corresponds to
$P_4$. The terms which have been retained suffice to determine the steady
states on the boundary and the linearization of the system at those points.
The point $P_5$ also appears in the second transformed system but since it
can be analysed in the chart corresponding to the first transformed system
the second transformed system, with $\bar y=\tilde y_2\tilde z_2$ and
$\bar z=\tilde z_2$, is only needed to analyse the steady state $P_6$ at the
origin of that coordinate system. For this purpose the only terms which need
to be retained are 
\begin{eqnarray}
&&{\tilde y_2}'=\tilde z_2[\gamma\alpha\tilde y_2+\cdots],
\label{trans2a}\\
&&{\tilde z_2}'=\tilde z_2[-\alpha(\gamma-1)\tilde z_2+\cdots].
\label{trans2b}
\end{eqnarray}
The common factor $\tilde z_2$ can be eliminated by a change of time coordinate.
The origin of this coordinate system corresponds to $P_6$. We see that in both 
cases, after a suitable change of time coordinate, the origin is a hyperbolic 
saddle. 

Next the centre manifold of $P_5$ will be studied in the case $\gamma=2$. We
do not expect that the case $\gamma>2$ is essentially different but since 
the algebra becomes significantly more complicated only the case $\gamma=2$
has been worked out. The centre subspace is parallel to the $\tilde y_1$-axis. 
We have ${\tilde z_1}'=O(\tilde y_1^3)$ on the centre manifold and this implies 
that if $\tilde z_1=\nu+w$ then 
\begin{equation}
2\alpha\nu w=[\nu^2(1-\nu)+2\alpha\nu^4+2\alpha\beta\nu^3]\tilde y_1^2+\ldots
\end{equation}
It follows that provided $\nu<1$ the flow on the centre manifold of $P_5$ is 
away from $P_5$. For the rest of the discussion we return to the case of
general $\gamma$.

In the case where $x$ gets large it remains to do one further quasihomogeneous 
directional blow-up of the origin in the $(Y,Z)$ coordinate system. In this 
case $(Y,Z)=(\bar y\bar z^{\gamma},\bar z^{\gamma-1})$. The time coordinate is
transformed using the relation $\frac{ds}{dt}=\frac{1}{\gamma-1}
\bar z^{\gamma^2-1}$. The resulting system is
\begin{eqnarray}
&&\bar y'=(\gamma-1)[\alpha\bar y^\gamma
-\alpha\bar y
-\alpha\nu\bar y^{\gamma+1}\bar z(1+\beta\bar z^{\gamma-1})\nonumber\\
&&+\bar y^{\gamma+1}\bar z^\gamma-\bar y\bar z^{\gamma-1}
-\nu\bar y^{\gamma+1}\bar z^\gamma(1+\beta\bar z^{\gamma-1})]\nonumber\\
&&-\gamma[\bar y^{\gamma+1}\bar z^\gamma
-\bar y\bar z^{\gamma-1}-\nu\bar y^{\gamma+1}\bar z^\gamma 
(1+\beta\bar z^{\gamma-1})],\\
&&\bar z'=\bar y^\gamma\bar z^{\gamma+1}
-\bar z^\gamma-\nu\bar y^\gamma\bar z^{\gamma+1} (1+\beta\bar z^{\gamma-1}).
\end{eqnarray}
There is a steady state at the point $(1,0)$ but since it is just another 
representation of $P_7$ it does not need to be analysed further. The origin
of this coordinate system corresponds to $P_8$. The 
$\bar z$-axis is a centre manifold for $P_8$ and the flow there is towards
$P_8$. Since the non-zero eigenvalue of the linearization at $P_8$ is negative
it can be concluded that $P_8$ is a sink.

Having completed the analysis of the case where $x$ gets large we now turn to
the case where where $y$ gets large (Case 2 in the terminology of 
\cite{brechmann18}), with new variables $X=\frac{x}{y}$ and $Z=\frac{1}{y}$.
The result is
\begin{eqnarray}
&&\frac{dX}{dT}=\frac{1}{Z^{\gamma+1}}[Z^{\gamma+2}-XZ+\nu Z(X+\beta Z)\nonumber\\
&&-\alpha X^2Z+\alpha XZ^{\gamma+1}+\alpha\nu X (X+\beta Z)],\\
&&\frac{dZ}{dT}=\frac{1}{Z^{\gamma+1}}
[-\alpha XZ^2+\alpha Z^{\gamma+2}+\alpha\nu Z(X+\beta Z)].
\end{eqnarray}
The common factor $\frac{1}{Z^{\gamma+1}}$ can be removed by a suitable change
of time coordinate satisfying $\frac{dt}{dT}=Z^{-\gamma-1}$. The linearization of
the resulting system about the origin
is identically zero so that it is again necessary to do a blow-up. In this
case a calculation using a Newton polygon gives the exponents $\alpha=1$ and
$\beta=1$. The transformation in the $X$ direction uses the relation
$(X,Z)=(\bar x_1,\bar x_1\bar z_1)$. The
resulting system is
\begin{eqnarray}
&&\frac{d \bar x_1}{dt}=\bar x_1[\bar x_1^{\gamma+1}\bar z_1^{\gamma+2}
-\bar x_1\bar z_1
+\nu\bar x_1\bar z_1 (1+\beta\bar z_1)\nonumber\\
&&-\alpha\bar x_1^2\bar z_1+\alpha\bar x_1^{\gamma+1}\bar z_1^{\gamma+1}
+\alpha\nu\bar x_1 (1+\beta\bar z_1)],\\
&&\frac{d \bar z_1}{dt}=\bar x_1[-\bar x_1^{\gamma}\bar z_1^{\gamma+3}+\bar z_1^2
-\nu\bar z_1^2 (1+\beta\bar z_1)].
\end{eqnarray}
The origin of this coordinate system corresponds to $P_3$. By a change of time
coordinate we can remove the factor $\bar x_1$. The linearization of the
system which results at the origin has one positive eigenvalue and the
$\bar z_1$-axis is invariant and defines a centre manifold at that point. It
can be concluded that $P_3$ is a source. If $\nu<1$ there is a steady state at
the point $\left(0,\frac{1-\nu}{\beta\nu}\right)$ which corresponds to the
point $P_2$. That point is a hyperbolic saddle whose stable manifold is the
$\bar z_1$-axis.

The transformation in the $Z$ direction uses the relation 
$(X,Z)=(\bar x_2\bar z_2,\bar z_2)$. The resulting system is
\begin{eqnarray}
&&\frac{d \bar x_2}{dt}=\bar z_2[\bar z_2^{\gamma}-\bar x_2
+\nu(\beta+\bar x_2)],\\
&&\frac{d \bar z_2}{dt}
=-\alpha XZ^2+\alpha Z^{\gamma+2}+\alpha\nu Z(X+\beta Z)\nonumber\\
&&=\bar z_2[-\alpha \bar x_2\bar z_2^2+\alpha\bar z_2^{\gamma+1}
+\alpha\nu\bar z_2(\beta+\bar x_2)].
\end{eqnarray}
The origin of this coordinate system is $P_1$.
By a change of time coordinate we can remove the factor $\bar z_2$. In the 
system which results there is inflow on the $\bar z_2$-axis while the 
$\bar x_2$-axis is invariant and corresponds to the $\bar z_1$-axis in the 
previous system. Note that the point $P_1$ is not a steady state.

The facts which have now been collected imply strong restrictions on the 
possible $\omega$-limit sets of solutions. The only points of the boundary
which they can contain are those on the part connecting $P_5$ and $P_7$. 
Poincar\'e-Bendixson theory implies that the $\omega$-limit set of a 
positive solution must be either a point (which can only be the positive
steady state, $P_7$ or $P_8$), a periodic solution or a heteroclinic cycle 
joining $P_5$ and $P_7$. The last of these can only occur if the centre 
manifolds of $P_5$ and $P_7$ coincide. Note that any periodic solution or
heteroclinic cycle must contain the positive steady state in its interior. 

We have the following analogue of Theorem 1 of \cite{brechmann18}.

\noindent
{\bf Theorem 2} There exists a positive number $\epsilon>0$ such that any
solution of the Michaelis-Menten system (\ref{selkovmm1})-(\ref{selkovmm2})
with initial data $x(0)=x_0$ and $y(0)=y_0$ which satisfies $x_0>\epsilon^{-1}$
and $x_0y_0^\gamma<\epsilon$ has the late-time asymptotics
\begin{eqnarray}
&&x(\tau)=\tau(1+o(1)),\\
&&y(\tau)=y_1e^{-\alpha\tau}(1+o(1)).
\end{eqnarray}
There exists a solution, unique up to time translation, which has the 
asymptotic behaviour
\begin{eqnarray}
&&x(\tau)=\tau(1+o(1)),\\
&&y(\tau)=\tau^{-\frac{1}{\gamma-1}}(1+o(1)).
\end{eqnarray}

\noindent
{\bf Proof} This theorem can be proved in the same way as Theorem 1 of
\cite{brechmann18}. The only extra element is that it is necessary to use the
fact that for this type of solution the time coordinates $\tau$ and $T$ are
asymptotically equal. The parameter $\nu$ does not contribute to the leading
order asymptotics. $\blacksquare$

\section{The global phase portrait}\label{portrait}

\begin{figure}
\centering
\begin{tikzpicture}[scale=1]
	\draw[->] (0,0)--(6,0);
	\draw[->] (0,0)--(0,4);
	\draw[dashed] (6,3)--(0,3) node[left]{\small $\frac{1}{\nu}$};
	\draw[dashed] (1.5,4)--(1.5,0) node[below]{\small $\frac{\beta \nu}{1- \nu}$};
	\draw[thick] (1.7,4) to [out=-88,in=100] (1.9,1.7) to [out=-80,in=170] (2.5,1) to [out=-10,in=177] (6,0.7);
	\draw[thick] (6,0.3) to [out=176,in=-10] (3.5,0.6) to [out=170,in=-90] (2.2,1.5) to [out=90,in=185] (5,2.7) to [out=5,in=183] (6,2.75);
	\node at (1.7,4) [right]{\small $N_1$};
	\node at (6,2.75) [below]{\small $N_2$};
	\draw[thick,->,gray] (2,0.7)--(2.3,0.5); 
	\node[gray] at (2.3,0.7) [below right]{\small $U_1$};
	\draw[thick,->,gray] (5,0.45)--(5.3,0.65);
	\node[gray] at (5.3,0.52) [right]{\small $U_2$};
	\draw[thick,->,gray] (4.7,1.7)--(4.5,1.9);
	\node[gray] at (4.7,1.8) [below right]{\small $U_3$};
	\draw[thick,->,gray] (2.4,2.6)--(2.2,2.4);
	\node[gray] at (2.4,2.7) [right]{\small $U_4$};
	\draw[fill] (2.39,1.03) circle(1pt);
\end{tikzpicture}
\caption{Nullclines.}
\label{Nullclines}
\end{figure}
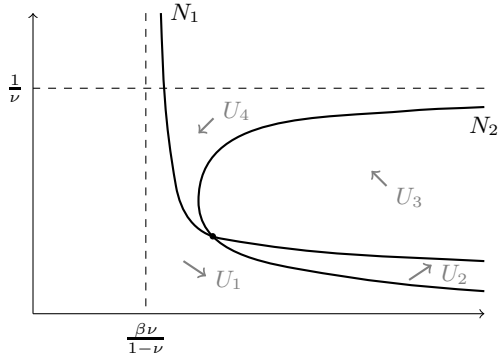

To understand the global phase portrait it is helpful to understand the 
geometry of the nullclines $N_1$ and $N_2$ given by $\dot x=0$ and $\dot y=0$,
respectively. We restrict consideration to the case $\nu<1$ where $N_1$ and
$N_2$ intersect in a single point. The equation for $N_1$ can be expressed
in the equivalent forms
\begin{equation}
y=\left[\frac{1}{-\beta\nu+(1-\nu)x}\right]^{\frac{1}{\gamma}},\ \ \
x=\frac{1}{1-\nu}(y^{-\gamma}+\beta\nu).
\end{equation}
Thus on $N_1$ the coordinate $y$ can be expressed as a smooth function of $x$
with a smooth inverse. Note, however, that while the second function is 
defined for all positive $y$ the first is only defined for 
$x>\frac{\beta\nu}{1-\nu}$. The equation for $N_2$ can be expressed in the form
\begin{equation}
x=\frac{y^{1-\gamma}+\beta\nu y}{1-\nu y}.
\end{equation}
Thus on $N_2$ the coordinate $x$ can be expressed as a (locally defined)
smooth function of $y$. Since $x$ can be written as a function of $y$ in both
cases and there is only one point of intersection it is clear that the 
complement of $N_1\cup N_2$ is a union of the four connected components defined
by the signs of $\dot x$ and $\dot y$. A schematic picture of the null clines
is given in Fig.~\ref{Nullclines}. As in \cite{brechmann18} we denote the regions with
the sign combinations $(+,-)$, $(+,+)$, $(-,+)$ and $(-,-)$ by $U_1$, $U_2$,
$U_3$ and $U_4$, respectively. Where $\dot y=0$ and $\dot x\ne 0$ we can use
the fact that $N_2$ is a graph over the $y$-axis to conclude that a solution
can only pass from $U_3$ to $U_4$ and $U_1$ to $U_2$ and not the other way 
round. Similarly the fact that $N_1$ is a graph over the $y$-axis implies 
that a solution can only pass from $U_4$ to $U_1$ and $U_2$ to $U_3$ and not 
the other way round. Thus the possible passages between the regions $U_i$ are 
just as in the case with mass action kinetics. Let $L$ be the part of the 
horizontal line segment joining the positive steady state to the $y$-axis
with the endpoint on the axis excluded. The part of $L$ excluding the steady
state is contained in $U_1$.  

\noindent
{\bf Lemma 7} In the Michaelis-Menten system each of the centre manifolds of
$P_5$ and $P_7$ contains a point of $L$ in its closure.

\noindent
{\bf Proof} A point on the centre manifold of $P_5$ which is sufficiently
close to $P_5$ lies in the region $U_3$. If we follow a solution which lies
on this manifold forwards in time then it must either tend to the positive
steady state as $t\to\infty$ or it must enter $U_4$ after a finite time and
in the latter case it must enter $U_1$. Once it has done so it must either
tend to the positive steady state as $t\to\infty$ or it must meet $L$
after a finite time. Similarly a solution on the centre manifold of $P_7$
which starts close to $P_7$ must, when followed backwards in time, either
converge to the positive steady state as $t\to -\infty$ or meet $L$
after a finite time. $\blacksquare$

Denote the $x$-coordinates of the points of $L$ in the closure of the centre 
manifolds of $P_5$ and $P_7$ for given value of $\alpha$ and $\nu$ by
$\xi_1(\alpha,\nu)$ and $\xi_2(\alpha,\nu)$.

\noindent
{\bf Lemma 8} The function $\xi_1-\xi_2$ describing the separation of the points
where the centre manifolds of $P_5$ and $P_7$ reach $y=1$ is continuous.

\noindent
{\bf Proof} The proof is similar to that of Lemma 2. The essential facts which
must be used are that any solution which approaches $P_5$ close enough in the
past time direction and does not lie on the centre manifold of $P_5$ cannot
remain close to $P_5$ forever and the corresponding statement with 'past'
replaced by 'future' and $P_5$ by $P_7$. $\blacksquare$

\noindent
{\bf Lemma 9} There are pairs of parameters $(\alpha,\nu)$ for which the
function $\xi_1-\xi_2$ describing the separation of the points where the
centre manifolds of $P_5$ and $P_7$ reach $y=1$ is negative. For $\nu>0$ fixed
and $0<\alpha\le\alpha_0$ either there exists an unstable periodic solution or
$\xi_1-\xi_2$ is positive. If there is some $\alpha\le\alpha_0$ for which no
periodic solutions exist then there exists an $\alpha_1$ with
$\xi_1(\alpha_1,\nu)= \xi_2(\alpha_1,\nu)$.

\noindent
{\bf Proof} For $\alpha$ sufficiently large the positive steady state is a
source and thus $\xi_1(\alpha,\nu)<1$. Thus in order to prove the first part
of the lemma it suffices to show that for some $(\alpha,\nu)$ we have
$\xi_2(\alpha,\nu)=1$. To prove this we proceed as in the case of mass action
kinetics. First the system is transformed to the coordinates $(\bar y,\bar z)$
and then the quantities $\epsilon$ and $\bar w$ are introduced. The right hand
side of each equation in the Michaelis-Menten case is the sum of the right
hand side of the corresponding equation in the mass action case and an
expression which can be written as $\epsilon^{-1}\nu$ times a function which is
regular in the limit $\epsilon\to 0$. Fixing $\nu$ and letting
$\epsilon$ tend to zero would cause this term to explode. Instead we let
$\epsilon$ and $\nu$ tend to zero in such a way that $\nu=\epsilon^2$. Then
the second summand behaves in a smooth manner as $\epsilon\to 0$ and in fact
tends to zero. Thus proceeding in the same way as in the proof of Lemma 3
gives the first conclusion. For the second part we can again proceed as in
the proof of Lemma 3. The difference is that while in the mass action case we
knew that there was no unstable periodic solution in the Michaelis-Menten
case we have to assume it. $\blacksquare$

Note that for the choice of parameters in the first part of Lemma 9 there is 
a heteroclinic orbit joining the positive steady state to the point $P_7$.
It follows that for these values of the parameters no periodic solutions exist.
This is because a periodic solution would have to contain the positive steady
state in its interior and therefore would have to cross the heteroclinic orbit.

There is no straightforward generalization of the monotonicity result
of Lemma 4 to the Michaelis-Menten case. The proof of monotonicity fails for
the centre manifold of $P_5$ since it may pass through the region $U_4$. For
this reason even in a case where the existence of a zero of $\xi_1-\xi_2$ can
be proved we do not get its uniqueness, Moreover, we do not get the analogue of
the stability statement in the mass action case.
It is possible to do a calculation analogous to that done to determine the
stability of the heteroclinic cycle in \cite{brechmann18}. Unfortunately in
the estimate for the return map the power $\gamma$ is replaced by the power one
and this gives no information about stability. The following theorem sums up
the results obtained.

\noindent
{\bf Theorem 3} The Michaelis-Menten system 
(\ref{selkovmm1})-(\ref{selkovmm2}) has the following properties. 

\noindent
1. For each choice of the parameters $(\alpha,\nu)$ with $\nu<1$ the unique
positive steady state has coordinates $\left(\frac{1+\beta\nu}{1-\nu},1\right)$.

\noindent
2. If $\gamma\le\frac{1+\beta\nu}{1-\nu}$ the steady state is stable. Otherwise
if $\alpha<\alpha_0=\frac{(1-\nu)^2}{\gamma (1-\nu)-(1+\beta\nu)}$ it is stable
and for $\alpha>\alpha_0$ unstable.

\noindent
3. For $\alpha=\alpha_0$ a generic Hopf bifurcation occurs. Parameters can be
chosen so as to make it supercritical or subcritical.

\noindent
4. For given $(\alpha,\nu)$ there exist positive numbers $x_0$ and $y_0$
such that if a solution satisfies $x(t)\ge x_0$ and $y(t)\le y_0$ at some time
$t$ then it has the late time asymptotics described in Theorem 2.

\noindent
5. For $\gamma=2$ there exists a choice of positive parameters $\alpha$ and 
$\nu$ for which all solutions other than the steady state have the late time 
asymptotics described in Theorem 2.

\noindent
6. If for $\gamma=2$ and given $\nu$ there exist no periodic solutions for 
$\alpha$ sufficiently small then there exits a heteroclinic cycle passing 
through the steady states $P_7$ and $P_5$ in that order.

It should be noted that it has not been shown here whether the case described in
point 6. ever occurs.

\section{Conclusions and outlook}

It has been shown that the basic Selkov system admits solutions with
unbounded oscillations and that the diameter of the image of a periodic
solution can tend to infinity as $\alpha$ approaches a finite limit, thus
completing the results of \cite{brechmann18} on that system and rigorously
confirming a claim made in \cite{selkov68}. Note that some statements related
to this issue have been made in \cite{erneux18} but that that reference does
not contain rigorous proofs of those statements. One remaining question is
that of the rate with which the diameter of the image of the periodic solution
tends to infinity as the critical parameter value $\alpha_1$ is approached. A
suggestion for this has been made in \cite{merkin87} for the case $\gamma=2$
but there is neither a rigorous proof that this suggestion is correct nor a
generalization of the statement to higher values of $\gamma$.

It was also investigated which properties of the basic Selkov system persist
in the Michaelis-Menten system from which Selkov derived his basic model.
Partial results were obtained and it was shown in particular that the
Michaelis-Menten system has unbounded solutions which are eventually monotone
for all parameter values. The question of whether the five-dimensional system
from which the Michaelis-Menten system itself was derived has unbounded
solutions remains open. It was shown that for suitable parameter values
unstable periodic solutions of the Michaelis-Menten system exist. It was left
open whether there exist unbounded oscillatory solutions or periodic solutions
whose images have arbitrarily large diameter for bounded ranges of the
parameters.

The unbounded solutions cast doubt on the suitability of the Selkov model for 
describing glycolytic oscillations. An alternative model often preferred to
the Selkov model is that of Goldbeter and Lefever \cite{goldbeter72}. There
the amplitude of the periodic solutions created in a Hopf bifurcation 
increases to a maximum before decreasing again to zero at a point where the
periodic solutions vanish again in a second Hopf bifurcation. It has been 
proved by d'Onofrio \cite{donofrio11}, on the basis of an analysis of a more 
general class of systems in \cite{othmer78}, that all solutions of the
Goldbeter-Lefever model are bounded. Further aspects of the dynamics of 
solutions of that model have been studied in \cite{donofrio11} and a 
sophisticated analysis of some of its properties has been carried out in
\cite{kosiuk11}.

The questions of the origin of the unbounded solutions and how they could be
eliminated by modifying the system have been discussed in \cite{merkin86}.
The origin of the unbounded growth can be seen in the constant source term
in the equation for $x$. This corresponds to an unlimited supply of the 
substrate. In \cite{merkin86} this is called the pooled chemical approximation.
If this is replaced by a mechanism where the substrate is formed
from a precursor which itself is limited in quantity then the oscillations
only grow within a finite time period before decaying again. An alternative 
modification is to introduce an additional uncatalysed conversion of the 
substrate into the product. The resulting system is called the (cubic)
autocatalator. According to the analysis of \cite{merkin86} this leads to a
situation similar to that described above for the Goldbeter-Lefever model and
the unbounded oscillations are absent. Some aspects of this type of model have 
been analysed rigorously in \cite{gucwa09}. 

The Selkov system and other related ones are model cases for understanding
oscillations in biological and chemical systems. The study of equations of
this type raises a number of issues. In what ways can heuristic and numerical 
results be made into rigorous theorems? How can we understand the relations
between the choices made in modelling and the relevance of the resulting
models for the applications? The present paper is intended as a contribution
to the clarification of these issues.

\vskip 10pt\noindent
{\it Acknowledgements} One of the authors (ADR) is grateful to James Sneyd and
J\'anos T\'oth for helpful discussions.


\begin{thebibliography}{12} 

\bibitem{alberts08} Alberts, B., Johnson, A., Lewis, J., Raff, M., Roberts, K.
and Walter, P. 2008 Molecular biology of the cell. Garland, New York.
\bibitem{boiteux75} Boiteux, A., Goldbeter, A. and Hess, B. 1975 Control of 
oscillating glycolysis of yeast by stochastic, periodic, and steady source of 
substrate: a model and experimental study. Proc. Natl. Acad. Sci. USA 72, 
3829–-3833.
\bibitem{brechmann18} Brechmann, P. and Rendall, A. D. 2018 Dynamics of the 
Selkov oscillator. Math. Biosci. 306, 152--159.
\bibitem{dumortier06} Dumortier, F., Llibre, J. and Art\'es, J. C. 2006 
Qualitative theory of planar differential systems. Springer, Berlin.
\bibitem{donofrio10} d'Onofrio, A. 2010 Uniqueness and global attractivity of
glycolytic oscillations suggested by Selkov's model. J. Math. Chem. 48, 
339--346.
\bibitem{donofrio11} d'Onofrio, A. 2011 Globally attractive oscillations in 
open monosubstrate allosteric enzymatic reactions. J. Math. Chem. 49, 531--545.
\bibitem{duysens57} Duysens, L. N. M. and Amesz, J. 1957 Fluorescence 
spectrophotometry of reduced phosphopyridine nucleotide in intact cells in the 
near-ultraviolet and visible region. Biochim. et Biophys. Acta 24, 19–-26.
\bibitem{erneux18} Erneux, T. 2018 Early models of chemical oscillations
failed to provide bounded solutions. Phil. Trans. R. Soc. A 376, 20170380.
\bibitem{goldbeter72} Goldbeter, A. and Lefever, R. 1972 Dissipative structures
for an allosteric model. Biophys. J. 12, 1302--1315.
\bibitem{gucwa09} Gucwa, I. and Szmolyan, P. 2009 Geometric singular 
perturbation analysis of an autocatalator model. Disc. Cont. Dyn. Sys. 
2, 783--806. 
\bibitem{higgins64} Higgins, J. 1964 A chemical mechanism for oscillation of
glycolytic intermediates in yeast cells. Proc. Natl. Acad. Sci. (USA) 51, 
989--994.
\bibitem{keener09} Keener, J. and Sneyd, J. 2009 Mathematical physiology. 
I: Cellular Physiology. Springer, Berlin.
\bibitem{kosiuk11} Kosiuk, I and Szmolyan, P. 2011 Scaling in singular 
perturbation problems: blowing up a relaxation oscillator. SIAM J. Appl. Dyn.
Sys. 10, 1307--1343.
\bibitem{kuehn15} Kuehn, C. 2015 Multiple time scale dynamics. Springer, Berlin.
\bibitem{merkin86} Merkin, J. H., Needham, D. J. and Scott, S. K. 1986 
Oscillatory chemical reactions in closed vessels. Proc. R. Soc. A406, 299-323.
\bibitem{merkin87} Merkin, J. H., Needham, D. J. and Scott, S. K. 1987 On the 
creation, growth and extinction of oscillatory solutions for a simple pooled
chemical reaction scheme. SIAM J. Appl. Math. 47, 1040--1060.
\bibitem{othmer78} Othmer, H. G. and Aldridge, J. A. 1978 The effects of cell
density and metabolite flux on cellular dynamics. J. Math. Biol. 5, 169-200. 
\bibitem{perko01} Perko, L. 2001 Differential Equations and Dynamical Systems.
Springer, Berlin.
\bibitem{selkov68} Selkov, E. E. 1968 Self-oscillations in glycolysis. I. A
simple kinetic model. Eur. J. Biochem. 4, 79--86.

\end{thebibliography}
\end{document}